\documentclass[aps,nofootinbib]{revtex4-2}
\usepackage{amssymb}
\usepackage{eurosym}
\usepackage{amsfonts}
\usepackage{geometry}
\usepackage[normalem]{ulem}
\usepackage{soul}
\usepackage{color}

\def\be{\begin{equation}}
\def\ee{\end{equation}}
\def\bea{\begin{eqnarray}}
\def\eea{\end{eqnarray}}
\begin{document}
\title{Quantum perturbative solutions of extended Snyder and Yang
models with spontaneous symmetry breaking}
\author{Jerzy Lukierski}
\affiliation{Institute of Theoretical Physics, Wroc\l aw
University, pl. Maxa Borna 9,
50-205 Wroc\l aw, Poland}
\author{Stjepan Meljanac}
\affiliation{
Division of Theoretical Physics, Rudj{}er
Bo\v{s}kovi\'c Institute, Bijeni\v{c}ka~c.54, HR-10002~Zagreb, Croatia}
\author{Salvatore Mignemi}
\affiliation{Dipartimento di Matematica, Universit\`a di
Cagliari via Ospedale 72, 09124 Cagliari, Italy and INFN, Sezione di
Cagliari, 09042 Monserrato, Italy}
\author{Anna Pacho\l}
\affiliation{Department of Microsystems, University of South-Eastern Norway, Campus Vestfold, Norway}
\begin{abstract}
We propose $\hbar$-expansions as perturbative solutions of quantum extended
Snyder and Yang models, with $\hbar$-independent classical zero-th order
terms responsible for the spontaneous breaking of $D=4$ and $D=5$ de Sitter
symmetries. In such models, with algebraic basis spanned by $\hat o(D,1)$
Lie algebra generators, we relate the vacuum expectation values (VEV) of the
spontaneously broken generators with the Abelian set of ten (Snyder, $D=4$) or
fifteen (Yang, $D=5$) antisymmetric tensorial generalized coordinates, which are also
used as zero order input for obtaining the perturbative solutions of quantum extended Snyder and Yang
models. In such a way we will attribute to these Abelian generalized coordinates the physical meaning of the order parameters describing spontaneous symmetry breaking (SSB).
It appears that the consecutive terms in $\hbar$-power series can be calculated explicitly if we supplement the SSB order parameters by the dual set of tensorial commutative momenta.
\end{abstract}
\maketitle
\section{Introduction}
Snyder and Yang models,  proposed in the first half of XX-th century \cite{Snyder:1946qz}, \cite{Yang}, were precursors of modern noncommutative geometry.
They are based on the idea of identifying the quantum space-time with the operators of a noncommutative algebra. In Snyder and Yang models the commutation relation between coordinates is proportional to the curvature of quantum positions \cite{Snyder:1946qz}, \cite{Yang} and Yang model also includes quantum noncommuting momenta \cite{Yang}. The quantum structure of space-time in these models permits to introduce non-trivial commutators between the components of quantum position and quantum momenta operators without explicitly breaking Lorentz
invariance. Such modifications of phase space commutation relations are expected in the algebraic description of quantum gravity (QG) and are important for investigating possible physical effects arising as quantum gravitational corrections.

The main aim of this paper is to present a novel approach to the perturbative solutions of extended Snyder and
Yang models and disclose in them the spontaneous symmetry breaking (SSB) effects. We  construct the operator-valued perturbative solutions expressed as a power series in the Planck constant $\hbar$,  and explain the role of classical, commutative parts of such solutions as providing SSB. The use of a perturbative $\hbar $-expansion of the solutions (see e.g.~\cite%
{Itzykson:1980rh,Brodsky:2010zk}) permits to distinguish the classical, commuting terms
obtained in the limit $\hbar =0$, from the remaining $\hbar $-dependent
quantum parts. When the quantum model is formulated in terms of
Lie algebra symmetry generators, it is known that the presence of their classical parts leads to
SSB effects (see e.g.~\cite{Goldstone:1962es}-\cite
{Brauner:2014aha}). In particular $D=4$ extended Snyder model \cite{Girelli:2010wi}-\cite{Meljanac:2022qoa}, with its algebraic formulation described by ten independent $\hat{o}(4,1)
$ symmetry generators ($D=4$ de Sitter algebra) fits very well in such a framework.

It is well known that in quantum theories one can
consider two ways of breaking symmetries. The first, explicit symmetry
breaking, leads to modified  basic symmetry properties of the algebraic structure in the
 quantum models under consideration, e.g. one obtains the modified action integrals, quantum equations of motion
etc. The second way, SSB, does not change the basic symmetries of algebraic structures, but provides the solutions as quantum states with broken symmetries.
In such a case the SSB effects have been considered in Quantum Mechanics (QM) and Quantum
Field Theory (QFT) models, and in particular in Standard Model (SM), which describes the theory of elementary particles
by the tools of QFT (see e.g. \cite{9a,9b}). We recall that in SM the suitable SSB of local gauge symmetries leads to the Higgs mechanism \cite{10,11} generating the mass parameters which are necessary for the comparison with experiment.

In this paper we consider the presence of SSB in the description of quantum space-times and quantum deformed phase spaces. By using the tools of noncommutative (NC) geometry (see e.g. \cite{12,13,14,15}), various NC models (see e.g. \cite{Snyder:1946qz,Yang,18,19,20})
describing $D=4$ quantum space-times, quantum deformed phase spaces, as well as quantum symmetry groups have been obtained.
We recall that the first NC models with preserved $D=4$ relativistic
covariance were introduced as early as in 1947 by Snyder \cite{Snyder:1946qz} and Yang \cite{Yang}. These models
and their generalizations were subsequently considered in numerous papers (see e.g. \cite{22,23,24,Girelli:2010wi,26,Meljanac:2021jdr,Meljanac:2021drl}), where however the appearance of  SSB effects had not been pointed out\footnote{It
should be recalled, however, that in \cite{26,Meljanac:2020fde,Meljanac:2021iyk,29} the Snyder-type models were
solved perturbatively as embedded in the canonical vectorial and tensorial
Heisenberg algebras \cite{30,31} but without introducing the SSB interpretation.}.
In this paper, we will show that the introduction  of explicit $\hbar$-dependence and the use of perturbation
theory described by $\hbar-$power series\footnote{For the discussion of $\hbar$-power expansions in quantum theories, see e.g. \cite{Brodsky:2010zk}.} permits to provide the SSB interpretation of the obtained results.

The plan of our paper is the following. In next Sec.~2 we present a short algebraic description of quantum Snyder model with two deformation parameters: the elementary length $l$, often identified with Planck length $l_P$, and  the universal Planck constant $\hbar$ characterizing quantum theories\footnote{If we follow the standard description of quantum theory defined by the passage from Poisson brackets to quantum commutators (see e.g. \cite{A}), the limit $\hbar\to 0$ of the quantum commutators describes the transition from quantum to classical theory.
In this paper we study the $\hbar\to0$ limit of perturbative $\hbar$ expansion as describing the Abelian SSB parameters. However, it should be mentioned that recently such prescription has been challenged. In particular in quantum gravity were studied models with quantum solutions which contain "side by side" both classical and quantum parts (see e.g. \cite{B,C,D}).}. The $\hbar$-dependent $D=4$ dS algebra basis of Snyder algebra can be treated as the relativistic dS extension from $\hat{o}(3)$ to $\hat{o}(4,1)$ of the $\hbar$-dependent nonrelativistic $D=3$ angular momentum algebra, which is well-known from basic textbooks on QM. Further, we describe in the presence of SSB, the reducible structure of Hilbert spaces of states with irreducible components labelled by the order parameters which characterize the spontaneously broken solutions. For simplicity, in Snyder model, we will consider the degeneracy of quantum states parametrized by the four-vector parameter $a_\mu$ describing the curved translations in the coset $\hat{o}(4,1)/\hat{o}(3,1)$. It appears that the vacuum state $||0\rangle\rangle$ in Snyder model is degenerate, given by the direct integral of irreducible vacua $|0;a_\mu\rangle$
 \footnote{The direct integrals of Hilbert spaces for reducible quantum fields satisfying Wightman axioms were first considered by Borchers \cite{E}; see also Haag \cite{F}, who considered the spontaneously broken quantum states in BCS model \cite{G} describing superconductivity.}.
In Sec.~3 we consider the $\hbar$-perturbative solution of spontaneously broken Snyder model with explicit formulae providing the first- and second-order terms.
In Sec.~4 we deal with quantum $D=4$ Yang model which is described algebraically by $D=5$ dS algebra. Very recently such models were studied and generalized (see \cite{H,I,J}) with the participation of the authors of the present paper. Further in Sec.~4 we consider the $\hbar$-perturbative solutions of Yang model and present explicitly the leading term linear in $\hbar$. Finally, Sec. 5 contains a short discussion of results and provides new suggestions about the continuation of present research and proposes the possible generalizations.
\section{Quantum $D=4$ Snyder model and spontaneously broken $D=4$ dS symmetries}
\textit{a) Algebraic description of quantum $D=4$ Snyder model} \\
The algebraic $D=4$ Snyder model is determined by the quantum NC space-time
position generators $\hat{x}_{\mu }$ and Lorentz-algebra generators $M_{\mu
\nu }$ satisfying the following relation\footnote{
For simplicity we put $c=1$, because our paper is not aimed at the
consideration of relativistic corrections, characterized by inverse powers
of $c$. We stress, however, that the quantum nature of the model considered here
is underlined by the explicit dependence on the
Planck constant $\hbar $, in agreement with the historic formulation of the
Snyder model \cite{Snyder:1946qz}.}
\begin{equation}
\lbrack \hat{x}_{\mu },\hat{x}_{\nu }]=i{\frac{l^{2}}{\hbar }}\hat{M}_{\mu
\nu },  \label{snyderx}
\end{equation}%
where $l$ is an elementary length, $\mu ,\nu =0,1,2,3$, and
\begin{equation}
\lbrack \hat{M}_{\mu \nu },\hat{M}_{\rho \tau }]=i\hbar (\eta _{\mu \rho
}\hat{M}_{\nu \tau }-\eta _{\mu \tau }\hat{M}_{\nu \rho }+\eta _{\nu \tau }\hat{M}_{\mu \rho
}-\eta _{\nu \rho }\hat{M}_{\mu \tau }),  \label{snyderMM}
\end{equation}%
\begin{equation}
\lbrack \hat{M}_{\mu \nu },\hat{x}_{\rho }]=i\hbar (\eta _{\mu \rho }\hat{x}%
_{\nu }-\eta _{\nu \rho }\hat{x}_{\mu }).  \label{snyderMx}
\end{equation}%
Relations (\ref{snyderMx}) show that Snyder quantum space-time
coordinates $\hat{x}_{\mu }$ describe a Lorentz-covariant four-vector. Using
the Compton length formula, $l={\frac{\hbar }{Mc}}$, after setting $c=1$ in
relation (\ref{snyderx}), we are led to the following form\footnote{
If $c=1$, the length $l$ and the mass $M$ are related by the "quantum" mass-length
relation $l=\hbar /M$, see e.g.~\cite{Brodsky:2010zk}.}
\begin{equation}  \label{xx}
\lbrack \hat{x}_{\mu },\hat{x}_{\nu }]=i{\frac{\hbar }{M^{2}}}\hat{M}_{\mu
\nu },  \label{snyderxx}
\end{equation}%
where $M$ is an elementary mass (e.g. Planck mass). If we introduce
\begin{equation}
\hat{M}_{4\mu }=M\hat{x}_{\mu },  \label{M4m}
\end{equation}
one can describe the relations (\ref{snyderMM})-(\ref{snyderxx}) as providing $D=4$ de
Sitter algebra
\begin{equation}
\lbrack \hat{M}_{AB},\hat{M}_{CD}]=i\hbar (\eta _{AC}\hat{M}_{BD}-\eta
_{AD}\hat{M}_{BC}+\eta _{BD}\hat{M}_{AC}-\eta _{BC}\hat{M}_{AD}),  \label{LorentzMM}
\end{equation}
with $A,B,=0,1,2,3,4$.

Originally, the Snyder model (\ref{snyderMM})-(\ref{snyderxx}) was introduced by
adding to Snyder quantum space-time $\hat{x}_{\mu }$ the
commuting four-momenta $p_{\mu }$, what leads to the
description of Snyder quantum phase space \cite{Snyder:1946qz}, \cite{Battisti:2010sr}. In such a case the generators $\hat{M}_{\mu \nu }$ can be
expressed in terms of the quantum phase space coordinates ($\hat{x}_{\mu }$,
$p_{\mu }$), which after the use of relations (\ref{snyderMM})-(\ref{snyderxx}) and Jacobi identities lead to the set of quantum-deformed Lorentz-covariant
Heisenberg algebras \cite{Battisti:2010sr,24,Mignemi:2013aua}.
{The structure constants of the algebra (\ref{LorentzMM}) are proportional to the Planck constant $\hbar$. Special realization of { algebras (\ref{snyderMM}) and} (\ref{LorentzMM}) in symmetric ordering can be written as power series in structure constants i.e.~power series in $\hbar$ \cite{30}.
All other realizations could be obtained using similarity transformations from special realization in symmetric ordering, for extended Snyder model see e.g.~\cite{Meljanac:2020fde}. In the limit $\hbar\to 0$, the {de Sitter} algebra (\ref{LorentzMM}) becomes an Abelian algebra.
In the limit $M\to\infty$  Snyder algebra (\ref{snyderxx}) reduces to $[ x_\mu, x_\nu]=0$, where $x_\mu$ are the commutative coordinates.}\\

\textit{b) Spontaneous breaking of $D=4$ dS symmetries}\\ In this paper we investigate the class of models with independent
generators $\hat{x}_{\mu }$, $\hat{M}_{\mu \nu }$ satisfying  eqs.~(\ref{snyderMM})-(\ref{snyderxx}), which define\footnote{We will consider the most physical $D=4$ case, but the results can be extended in a straightforward way to any dimension $D\geq 2$, with the Snyder algebras spanned by generators of $\hat{o}(D,1)$.} a ten-dimensional independent algebraic basis  of $D=4$ Snyder model. Such models were studied during the last twenty years and were
named alternative \cite{Girelli:2010wi} or extended \cite
{Meljanac:2021jdr,Meljanac:2020fde,Meljanac:2021iyk,Meljanac:2021drl,29,Meljanac:2022qoa}. They were solved perturbatively in terms of tensorial
canonical quantum phase space coordinates $(x_{AB};p_{AB})\equiv (x_{\mu \nu }, {M x_{\mu
}};p_{\mu \nu }, p_{\mu })$, where\footnote{
In some of our papers (see e.g. \cite{Meljanac:2020fde}, \cite{Meljanac:2021drl}, \cite{30}) we used the tensorial canonical Heisenberg algebras, but we did not consider their explicit $\hbar$-dependence. In most of our earlier papers, related with Snyder models, \cite
{Meljanac:2021jdr,Meljanac:2020fde,Meljanac:2021iyk,Meljanac:2021drl,29,Meljanac:2022qoa} we considered the Heisenberg algebra relations (\ref{5dalgebra}) with $\hbar=1$.}
\begin{equation}
\lbrack x_{AB},x_{CD}]=[p_{AB},p_{CD}]=0,\qquad \lbrack
x_{AB},p_{CD}]=i\hbar (\eta _{AC}\eta _{BD}-\eta _{AD}\eta _{BC}).
\label{5dalgebra}
\end{equation}
The novelty of this paper is to express the generators $\hat{M}_{AB}$ as an algebraic $\hbar $-power series and consider the zero-th order terms $\hat{M}_{AB}^{(0)}=x_{AB}$ as representing Nambu-Goldstone (NG) modes which describe the spontaneous symmetry breaking of $D=4$ de Sitter symmetry. One can study
the following two particular choices:

i) $x_{\mu }\neq 0$, $x_{\mu \nu }=0$. In such a case the Lorentz symmetry
is not broken, and the NG modes are determined by the parametrization of the coset $\hat{o}(4,1)/\hat{o}(3,1)$, which  describes the curved de Sitter translations (see e.g. \cite{Zumino:1977av}).

ii) $x_{\mu }=0$, $x_{\mu \nu }\neq 0$. This case corresponds to
spontaneously broken Lorentz symmetries (see e.g.~\cite{Volkov}, \cite{Bluhm:2004ep}, \cite{Brauner:2014aha}, \cite{Kostelecky:2009zr}).

The canonical coordinates $x_{AB}$ (see (\ref{5dalgebra})) are given by the classical $\hbar $-independent part of the $\hbar $-expansions,
\begin{equation}\label{xM_cl}
\hat{x}_{\mu }=x_{\mu }^{(cl)}+\hat{x}_{\mu }^{(q)},\qquad \hat{M}_{\mu \nu
}=x_{\mu \nu }^{(cl)}+\hat{x}_{\mu \nu }^{(q)},
\end{equation}
where $x_{\mu }$ and $x_{\mu \nu }$
describe the zero-th order in the $\hbar $-power expansions and describe the classical parts of the generators $\hat{M}_{AB}$.
In quantum models with preserved $D=4$ dS symmetries the classical parts of $\hbar$-power series vanish and one can introduce a
unique (invariant under symmetries) cyclic vacuum state
$|0 \rangle$ ($\langle 0\,|\,0\rangle =1$), which satisfies the relations
\begin{equation}
\hat{x}_{\mu }^{(q)}|0\rangle=0,\qquad \hat{x}_{\mu \nu }^{(q)}|0\rangle=0 . \label{vac}
\end{equation}
In general case, if $x_{AB}\neq 0$, one should introduce the degenerate continuous set of vacua $|0;x_{AB}\rangle$. The commuting
coordinates $x_{AB}$ define the order parameters which
describe spontaneously broken rotations in the planes $(A,B)$\footnote{If $A=0$ the rotational symmetry is $\hat{o}(1,1)$, if $A=1,\ldots, 4$ we deal with SSB of $\hat{o}(2)$ rotations.} which are the Abelian subgroups of spontaneously broken $D=4$
de Sitter symmetry.

%
%
Let us discuss the $D=4$ Snyder model with spontaneous symmetry breaking, which is generated by the curved $D=4$ dS translations, parametrized by a constant four-vector $a_\mu$ (see e.g. \cite{Zumino:1977av,Banerjee})\footnote{In \cite{Zumino:1977av,Banerjee}, the analogous case of nonlinear curved $D=4$ AdS translations is considered.}.
In such a case the SSB is generated by the action of nonlinear unitary representation $U(a_\mu)$ on the NC curved space-time coordinates $\hat{x}_\mu$, which results in the following inhomogeneous nonlinear formulae \cite{Zumino:1977av,Banerjee}:
\begin{equation}\label{a}
\hat{x}{(a_\mu)}_\mu =U^{-1}(a_\mu)\hat{x}_\mu U(a_\mu)=\hat{x}_\mu+ a_\mu +O(\hat{x}_\mu;a_\mu)
\end{equation}
where $O(\hat{x}_\mu;a_\mu)$ contains higher powers of $\hat{x}_\mu$.
In such a case one can introduce the continuous sets of degenerate vacuum states $|0;a_\mu\rangle$ and the $a_\mu$-dependent Hilbert spaces $\mathcal{H}(a_\mu)$, with different values of $a_\mu$ linked by the unitary representation $U(a_\mu)$ as follows:
\begin{equation}\label{c}
|0;a_\mu\rangle\rightarrow |0;a_\mu + a_\mu '\rangle = U(a_\mu ')|0;a_\mu\rangle
\end{equation}
and
\begin{equation}\label{b}
\mathcal{H}(a_\mu)\rightarrow \mathcal{H}(a_\mu + a_\mu ') = U(a_\mu ')\mathcal{H}(a_\mu),
\end{equation}
i.e. to each Hilbert space $\mathcal{H}(a_\mu)$ there exists an associated spontaneously broken set of vacua states (\ref{c}).
The total Hilbert space $\mathcal{H}$ describing whole spontaneously broken  quantum system can be described by the direct integral of Hilbert spaces $\mathcal{H}(a_\mu)$\footnote{The direct integrals of Hilbert spaces and degenerated vacua for reducible quantum fields satisfying Wightman axioms were considered in \cite{E}. In \cite{F} was considered the quantum BCS model \cite{Bardeen}, and in \cite{Lopuszanski} was studied a toy model of QFT with degenerate vacuum and reducible Hilbert spaces.}, with all possible values of the numerical parameters $a_\mu$ described by a classical manifold $V$ with Lebesgue measure $d\mu$
\begin{equation}\label{d}
    \mathcal{H}=\int^{\oplus a_\nu}_V d\mu \mathcal{H}(a_\nu).
\end{equation}
Analogously, the reducible degenerated vacuum $||0\rangle\rangle$ covariant under the spontaneously broken curved translation symmetries can be defined by the formula
\begin{equation}\label{d}
    ||0\rangle\rangle=\int_V^{\oplus a_\mu}d\mu |0;a_\mu\rangle.
\end{equation}

\section{$\hbar$-perturbative solutions of spontaneously broken $D=4$ Snyder model}
We firstly apply the scheme of perturbative $\hbar $-expansions to the
extended Snyder model (see, e.g. (\ref{snyderx}-\ref{M4m})), with the
algebra described by $\hat{o}\left( 4,1\right)$ generators $\hat{M}_{AB}=(%
\hat{M} _{\mu \nu },\hat{M}_{4\mu }=M\hat{x}_{\mu })$. We expand the
generators $\hat{M}_{AB}$ in the following $\hbar $-power series:
\begin{equation}\label{Mseries}
\hat{M}_{AB}=M_{AB}^{\left( 0\right) }+\hbar \hat{M}_{AB}^{\left( 1\right)
}+\hbar ^{2}\hat{M}_{AB}^{\left( 2\right) }+...
\end{equation}
where
\begin{equation}
M_{AB}^{\left( 0\right) }\equiv M_{AB}^{(cl)}=\lim_{\hbar \rightarrow 0}\hat{%
M}_{AB}=\langle\langle0||\hat{M}_{AB}||0\rangle\rangle
\end{equation}
or equivalently (see (\ref{vac}))
\begin{equation}
M_{AB}^{\left( 0\right) }\equiv x_{AB}=\left( x_{\mu \nu },{M}x_{\mu }\right)
\end{equation}
%
where $x_{AB}$ are the order parameters describing the SSB of the ten one-dimensional (pseudo-) orthogonal symmetries generated by $\hat{M}_{AB}$ on all planes $(A,B)$, where $(A,B=0,1,2,3,4)$, of $D=5$ space-time with signature $\eta_{AB}=diag (-1,1,1,1,1)$).

   One can deduce from the relations (\ref{snyderx})-(\ref{M4m}) the iterated set of algebraic equations determining the perturbative quantum terms $\hat{M}^{(n)}_{AB}$ $(n= 1,2,3,...)$ as functions of $x_{AB}$, describing Nambu-Goldstone (NG) degrees of freedom $x_{AB}$ and dual momenta $p_{AB}$, which satisfy together the generalized canonical quantum phase space relations (\ref{5dalgebra}). The most general case, when all $x_{AB}\neq0$, describes the situation when all the $D=4$ deS symmetries are spontaneously broken.
%
\smallskip
\newline
\textit{a) Perturbative $\hbar $-expansion: first order in $\hbar $}\\
\indent From relation (\ref{LorentzMM}) one gets:
\begin{equation}  \label{xAB_MCD}
\left[ x_{AB}{},\hat{M}_{CD}{}^{\left( 1\right) }\right] +\left[ \hat{M}%
_{AB}{}^{\left( 1\right) },x_{CD}{}\right] =i(\eta _{AC}x_{BD}-\eta
_{AD}x_{BC}-\eta _{BC}x_{AD}+\eta _{BD}x_{AC})
\end{equation}
 and relations (\ref{snyderMx},\ref{xx})
lead to
\begin{equation}  \label{xmn_Mrs}
\left[ x_{\mu\nu}{},\hat{M}_{\rho\sigma}{}^{\left( 1\right) }\right] -\left[ x_{\rho\sigma}{},\hat{M}
_{\mu\nu}{}^{\left( 1\right) }\right] =i(\eta _{\mu\rho}x_{\nu\sigma}-\eta
_{\mu\sigma}x_{\nu\rho}-\eta _{\nu\rho}x_{\mu\sigma}+\eta _{\nu\sigma}x_{\mu\rho}),
\end{equation}
\begin{equation}  \label{S_firstxM}
\left[ x_{\mu }{},\hat{M}_{\rho \sigma}{}^{\left( 1\right) }\right] -\left[
x_{\rho \sigma }{},\hat{x}_{\mu }{}^{\left( 1\right) }\right] =i(\eta _{\mu
\sigma }x_{\rho }-\eta _{\mu \rho }x_{\sigma }),\quad
\end{equation}
\begin{equation}  \label{S_firstxx}
\left[ x_{\mu }{},\hat{x}_{\nu }{}^{\left( 1\right) }\right] -\left[ x_{\nu
},\hat{x}_{\mu }{}^{\left( 1\right) }{}\right] =\frac{i}{M^{2}}x_{\mu \nu
}.\quad
\end{equation}
In order to solve the relations (\ref{xAB_MCD})-(\ref{S_firstxx}) we employ the generalized momenta $p_{AB}=\left(p_{\mu \nu },p_{\mu }\right) $
(see (\ref{5dalgebra})). From (\ref{xmn_Mrs}) and (\ref{S_firstxM}) one
can obtain a  particular solution, given by
\be\label{solM1}
\hbar\hat{M}_{\mu \nu ;S}^{\left( 1\right) }
=\frac{1}{2}\left(x_{\mu }^{~\ \rho }p_{\nu\rho }-x_{\nu }^{~\ \rho }p_{\mu\rho}\right)+
x_{\mu }p_{\nu }-x_\nu p_\mu
\ee
and in consistency with (\ref{S_firstxx})
\be\label{solx1}
\hbar\hat{x}_{\mu ;S}^{\left( 1\right) }=-\frac{1}{2M^{2}}x_{\mu \rho }p^{\rho }.
\ee
The general first order solution depends on one free parameter \cite{Meljanac:2020fde}
and can be obtained by a suitable choice of
 similarity transformations of the particular solutions (\ref{solM1},\ref{solx1}).
 \smallskip
  \newline
\textit{b) Perturbative $\hbar $-expansion: second order in $\hbar $} \\
\indent The second order counterpart of relation (\ref{xAB_MCD}) looks as follows:
\begin{equation}
\left[ x_{AB},\hat{M}_{CD}^{\left( 2\right) }\right] -\left[ x_{CD},\hat{M}%
_{AB}^{\left( 2\right) }\right] +\left[ \hat{M}_{AB}^{\left( 1\right) },\hat{%
M}_{CD}^{\left( 1\right) }\right] =i(\eta _{AC}\hat{M}_{BD}^{\left( 1\right)
}+\eta _{BD}\hat{M}_{AC}^{\left( 1\right) }-\eta _{BC}\hat{M}_{AD}^{\left(
1\right) }-\eta _{AD}\hat{M}_{BC}^{\left( 1\right) })
\end{equation}
which leads to:
\be\label{sec_or1}
\left[ x_{\mu \nu },\hat{M}_{\rho \sigma }^{\left( 2\right) }\right] -\left[
x_{\rho \sigma },\hat{M}_{\mu \nu }^{\left( 2\right) }\right] =i(\eta _{\mu
\rho }\hat{M}_{\nu \sigma }^{\left( 1\right) }+\eta _{\nu \sigma }\hat{M}_{\mu
\rho }^{\left( 1\right) }-\eta _{\nu \rho }\hat{M}_{\mu \sigma }^{\left(
1\right) }-\eta _{\mu \sigma }\hat{M}_{\nu \rho }^{\left( 1\right) })-\left[
\hat{M}_{\mu \nu }^{\left( 1\right) },\hat{M}_{\rho \sigma }^{\left( 1\right) }
\right],
\ee
\be\label{sec_or2}
\left[ x_{\mu },\hat{M}_{\rho \sigma }^{\left( 2\right) }\right] -\left[
x_{\rho \sigma },\hat{x}_{\mu }^{\left( 2\right) }\right] =i(\eta _{\mu \sigma }\hat{x}_\rho^{(1)}-\eta _{\mu \rho}\hat{x}_\sigma^{(1)})-\left[ \hat{x}_{\mu }^{\left( 1\right)
},\hat{M}_{\rho \sigma }^{\left( 1\right) }\right],
\ee
\be\label{sec_or3}
\left[ x_{\mu },\hat{x}_{\sigma }^{\left( 2\right) }\right] -\left[ x_{\sigma },
\hat{x}_{\mu }^{\left( 2\right) }\right] =\frac{i}{M^{2}}\hat{M}_{\mu \sigma
}^{\left( 1\right) }-\left[ \hat{x}_{\mu }^{\left( 1\right) },\hat{x}_{\sigma
}^{\left( 1\right) }\right].
\ee 
Substituting in (\ref{sec_or1})-(\ref{sec_or3}) the solutions (\ref{solM1}),(\ref{solx1}) one gets the particular solution, to second order in $\hbar$\footnote{The subscript $S$ denotes
the Snyder case. The factor $\hbar$ on the left hand side in (\ref{solM1}),(\ref{solx1}) and $\hbar^2$ in (\ref{solM2}),(\ref{solx2}) reflect the property that we deal with quantum-mechanical momenta satisfying the relations (\ref{5dalgebra}), proportional to $\hbar$ (one can recall the space-time realization $p_\mu=-i\hbar\partial_\mu$). Relation (\ref{solM1})
describes generalized angular momentum, in space-time realization, given by the $\hbar$-independent formula $M^{(1)}_{\mu\nu}=i\left(x_{[\mu}\partial_{\nu]}+\frac{1}{2}x_{[\mu}\,^\rho\partial_{\nu]\rho}\right)$. In the general case, the coefficients $M^{(n)}_{AB}$
in (\ref{Mseries}) are proportional to $n$-th powers of the canonical momenta (\ref{5dalgebra}) and are $\hbar$-independent.}:
\begin{equation}
\hbar^2 \hat{M}_{\mu \nu ;S}^{\left( 2\right) }=-\frac{1}{12}\left(x_{\mu \rho }
p^{\rho \sigma }p_{\nu \sigma }-x_{\nu \rho }p^{\rho \sigma }p_{\mu \sigma }
-2x^{\rho \sigma }p_{\mu \rho }p_{\nu \sigma }\right),
\label{solM2}
\end{equation}%
\begin{equation}
\hbar^2 \hat{x}_{\mu ;S}^{\left( 2\right) }=\frac{1}{M^2}\left(x_\rho p^\rho p_\mu +\frac{1}{4}\left( x_{\mu \rho }p_{\rho \sigma }
p_\sigma +x^{\rho \sigma }p_\rho p_{\mu \sigma }\right)\right).
\label{solx2}
\end{equation}%
General solutions in the second $\hbar$-order can be obtained from the formulae (\ref{solM2}),(\ref{solx2}) by performing suitable similarity transformations.
One can also show that, in the perturbative $n$-th order in $\hbar$, the solutions $(\hat{x}_{\mu ;S}^{(n)}, \hat{M}_{\mu\nu ;S}^{(n)})$ are $n$-linear in momenta $p_\mu, p_{\mu\nu}$ (see also \cite{Meljanac:2020fde}, \cite{Meljanac:2021drl}).

\section{Quantum $D=4$ Yang model and spontaneously broken algebra $\hat{o}\left( 5,1\right)$}
In the following we will apply our method to $D=4$ Yang model (see e.g. \cite{Yang},\cite{Guo, Heckman,Manolakos}), algebraically described by fifteen generators of $D=5$ dS algebra $\hat{o}(5,1)$ ($K,L=0,1,2,3,4,5$)
\begin{equation}
\hat{M}_{KL}=\left( \hat{M}_{\mu \nu },\hat{M}_{4\mu }=M\hat{x}_{\mu },\hat{M%
}_{5\mu}=R\hat{q}_{\mu },\hat{M}_{45}=MR\hat{r}\right) \label{MKL}
\end{equation}
satisfying the following relation
\begin{equation}
\label{MKL}
\lbrack \hat{M}_{KL},\hat{M}_{PR}]=i\hbar (\eta _{KP}\hat{M}_{LR}-\eta
_{KR}\hat{M}_{LP}+\eta _{LR}\hat{M}_{KP}-\eta _{LP}\hat{M}_{KR}).
\end{equation}
The $D=4$ Yang model describes a $D=4$ Lorentz-covariant quantum-deformed relativistic Heisenberg algebra with two deformation
parameters $\left( M,R\right)$ of length dimensions $\left[ M\right]
=L^{-1},\left[ R\right] =L$ and one  dimensionless scalar Abelian $\hat{o}\left( 2\right) $ generator $\hat{r}$.
 In the general case one can introduce in the Yang model
fifteen Abelian NG modes $x_{KL}=-x_{LK}$, which break spontaneously the $\hat{o}\left( 5,1\right) $ symmetry{
\begin{equation}
x_{KL}=\left( x_{\mu \nu },Mx_{\mu },Rq_{\mu },MRr\right) .  \label{var_x}
\end{equation}}
In order to solve the Yang model by using a perturbative $\hbar $-expansion
one should introduce fifteen canonically conjugated commuting NG momenta
\begin{equation}
p_{KL}=\left( p_{\mu \nu },p_{\mu },k_{\mu },s\right).   \label{var_p}
\end{equation}
The variables (\ref{var_x}), (\ref{var_p}) satisfy $D=5$ extension of the canonical commutation relations $(\ref{5dalgebra})$, with the following Lorentz-covariant additional relations
\begin{equation}
\left[ q_{\mu },k_{\nu }\right] =i\hbar \eta _{\mu \nu },\quad \quad \left[
r,s\right] =i\hbar .  \label{qk}
\end{equation}
Using the variables (\ref{var_x}),(\ref{var_p}) we present below the first order $\hbar$-perturbative solution of the Yang model.
\smallskip
\newline
\textit{a) Algebraic description of $D=4$ Yang model}

The Yang model was obtained in \cite{Yang} as a group-theoretic extension by momentum sector of the extended Snyder model.
Such an extension can be obtained by the Born map applied to the Snyder model generators $\hat{x}_\mu\rightarrow\hat{p}_\mu$, $\hat{M}_{\mu\nu}\rightarrow\hat{M}_{\mu\nu}$ and adding the Born map-invariant scalar generator $\hat{r}$.

In the Yang model we extend the relations (\ref{snyderMM})-(\ref{snyderxx}) by
the following set of algebraic equations\footnote{In Yang model we denote curved noncommutative momenta by $\hat{q}_{\mu }$, while $\hat{q}_{\mu }^{(0)}=q_\mu$ describes their classical commutative limit. The canonically dual coordinates are $k_\mu$ (see (\ref{qk})) which are different from $x_\mu$. Obviously, we assume that $[\hat r, \hat M_{\mu\nu}] = 0$.}
\begin{eqnarray}  \label{qq}
\left[ \hat{q}_{\mu },\hat{q}_{\nu }{}\right] &=&i\frac{\hbar }{R^{2}}\hat{M}%
_{\mu \nu }, \\
\left[ \hat{x}_{\mu }{},\hat{q}_{\nu }\right] &=&i\frac{\hbar }{MR}\eta
_{\mu \nu }\hat{M}_{45},\quad \hat{M}_{45}=MR\cdot \hat{r}, \\
\left[ \hat{M}_{\mu \nu },\hat{q}_{\rho }{}\right] &=&i\hbar \left( \eta _{\mu \rho }\hat{q}_{\nu }-\eta
_{\nu \rho }\hat{q}_{\mu }\right) , \\
\left[ \hat{r},\hat{x}_{\mu }{}\right] &=& \frac{i\hbar}{M^2}\hat{q}_{\mu },
\\
\left[ \hat{r},\hat{q}_{\mu }{}\right] &=&-\frac{i\hbar}{R^2}\hat{x}_{\mu }
\label{rqhat}
\end{eqnarray}
where $\hat{q}_{\mu }=q_{\mu }^{(cl)}+\hat{q}_{\mu }^{(q)},\quad \hat{r}=r^{(cl)}+\hat{r}^{(q)}$. It should be added that in the Yang model the original phase space variables $(\hat{x}_\mu,\hat{q}_\mu)$ represent the generalized set of quantum coordinates, which can be doubled by Hopf-algebraic duality relations $(\hat{x}_\mu,\hat{q}_\mu\rightarrow \hat{x}_\mu,\hat{q}_\mu;\hat{p}_\mu,\hat{k}_\mu)$.
{In the limit $R\to\infty$, the Yang model becomes the Snyder model, while for  $M\to\infty$, we obtain the inhomogeneous $D=4$ de Sitter algebra in momentum space. When both $M\to\infty$ and $R\to\infty$ Yang model gives rise to the semidirect product of Poincar\'e algebra and  commutative four-momenta, supplemented by a scalar variable.}
\smallskip
\newline
\indent\textit{b) $\hbar$-perturbative expansion of Yang model - linear terms}\\
We obtain the first order $\hbar $-approximation of the algebraic solutions of the Yang model if in the $\hbar$-expansions of the solutions (\ref{snyderMM})-(\ref{xx}) and (\ref{qq})-(\ref{rqhat}) we consider the linear $\hbar $-terms. Besides (\ref{xmn_Mrs}-\ref{S_firstxx}) one gets
\begin{eqnarray}  \label{qq0}
\left[ q_{\mu }{},\hat{q}_{\nu }{}^{\left( 1\right) }\right] -\left[ q_{\nu
}{},\hat{q}_{\mu }{}^{\left( 1\right) }\right] &=&\frac{i}{R^{2}}x_{\mu \nu
}, \\
\left[ x_{\mu }{},\hat{q}_{\nu }{}^{\left( 1\right) }\right] -\left[ q_{\nu
}{},\hat{x}_{\mu }{}^{\left( 1\right) }\right] &=&ir\eta_{\mu\nu},\quad \left( r\equiv
\hat{r}^{\left( 0\right) }\right) ,  \label{xq} \\
\left[ x_{\mu \nu }{},\hat{q}_{\rho }{}^{\left( 1\right) }\right] +\left[
\hat{M}_{\mu \nu }{}^{\left( 1\right) },q_{\rho }{}\right] &=&i\left( \eta _{\mu \rho }q_{\nu }-\eta
_{\nu \rho }q_{\mu }\right) ,  \label{xrhoM} \\
\left[ r{},\hat{x}_{\mu }{}^{\left( 1\right) }\right] +\left[ \hat{r}%
^{\left( 1\right) },x_{\mu }\right] &=&\frac{i}{M^2}q_{\mu },  \label{rx} \\
\left[ r{},\hat{q}_{\mu }{}^{\left( 1\right) }\right] +\left[ \hat{r}%
^{\left( 1\right) },q_{\mu }\right] &=&-\frac{i}{R^2}x_{\mu }.  \label{rq}
\end{eqnarray}
For the extended Snyder model, in the first order, we obtained the formulas (\ref{solM1}),(\ref{solx1}). In Yang model, due to the presence of additional
coordinates $(q_\mu,r)$ and momenta $(k_\mu,s)$, see (\ref{var_x}),(\ref{var_p}), one should extend the formulae (\ref{solM1}),(\ref{solx1}) by terms which are linear in momenta $(k_\mu,s)$, see (\ref{var_p}). We get ($a,b,c,d$ are numerical constants):
\be
\hbar\hat{M}_{\mu \nu ;Y}^{\left( 1\right) }=\frac{1}{2}\left(x_{\mu }^{~\ \rho }p_{\nu\rho }-x_{\nu }^{~\ \rho }p_{\mu\rho}\right)+
x_{\mu }p_{\nu }-x_\nu p_\mu {-} q_{\mu }k_{\nu } {+} q_{\nu }k_{\mu } ,
\ee
\be\label{sol_x1Y}
\hbar\hat{x}_{\mu ;Y}^{\left( 1\right) }=-\frac{1}{2M^2}x_{\mu\rho}p^\rho +ax_{\mu\rho}k^{\rho}+brk_\mu+cq_\mu s,
\ee
and add the following formulae for the first order solutions of $\hat{q}_\mu$ and $\hat{r}$:
\be\label{sol_q1Y}
\hbar\hat{q}_{\mu }{}^{\left( 1\right) }=-\frac{1}{2R^{2}}x_{\mu \rho }{k}^{\rho }+\tilde{a}x_{\mu \rho }p^{\rho }+\tilde{b}rp_{\mu }+\tilde{c}x_\mu s,
\ee
\be\label{sol_r}
\hbar\hat{r}{}^{\left( 1\right) }=dq^\rho p_{\rho }+fx^\rho k_\rho,
\ee
which depend on five additional numerical constants $\tilde{a},\tilde{b},\tilde{c},d$ and $f$. The equations (\ref{qq0})-(\ref{rq}) impose the following constraints on the eight parameters in (\ref{sol_x1Y})-(\ref{sol_r}):
\be
a+\tilde{a}=0,\qquad \tilde{b}=b+1, \qquad c-d=\frac{1}{M^2}, \qquad \tilde{c}-f=-\frac{1}{R^2}
\ee
and, in formulae (\ref{sol_x1Y})-(\ref{sol_r}), imply the absence of terms proportional to $p_{\mu\nu}$. We see therefore that the solutions of equations (\ref{qq0})-(\ref{rq}) which are linear in $\hbar$ contain four unconstrained numerical parameters $a,b,c,f$.

The above calculation can be extended to higher orders in $\hbar$, what we plan to present in a forthcoming publication.
\section{Outlook and final remarks}
The basic idea of Snyder and Yang models relies on the use of $D=4$ and $D=5$ de Sitter algebras for the algebraic description of, respectively, relativistic noncommutative quantum space-times and quantum phase spaces with noncommutative four-momenta.
In this paper, the quantum nature of Snyder and Yang models has been underlined by
considering their explicit dependence on the Planck constant $\hbar$, in agreement with the first historical formulations of both models \cite{Snyder:1946qz}, \cite{Yang}. By using $\hbar$ as an expansion parameter in the perturbative solutions, we were able to interpret the generalized tensorial coordinates,
introduced in our earlier papers (see e.g. \cite{Meljanac:2020fde}-\cite
{Meljanac:2022qoa}, \cite{30}) and we present them here as appearing due to the effects of
spontaneous symmetry breaking of $D=4$ and $D=5$ de Sitter symmetries.
We should also add that Snyder and Yang models can be considered as providing examples of the noncommutative space-times and quantum deformed phase spaces  which are considered in quantum gravity studies as the physics related applications of noncommutative geometry.

In our future work we plan to study the generalizations and
modifications of the models considered here, in particular:\\
\indent i) The $\kappa$-deformed extended Snyder models (see \cite{Meljanac:2021jdr}, \cite{Meljanac:2021drl}, \cite{Meljanac:2011mt}) were obtained by adding to the basic
deformation parameter $M$ the second parameter $\kappa$ with mass-like
dimension, in a way which leads in the limit $M\rightarrow\infty$ to the
well-known $\kappa$-deformed quantum Minkowski space-time (see e.g. \cite{MR}, \cite{Lukierski}).
Similarly in Yang model with the pair of basic deformation parameters $M$ and $R$ (see Sec.~4) one can add a pair of parameters ($\kappa$, $\tilde{\kappa}$) and introduce doubly $\kappa$-deformed Yang models  with $\kappa$-deformed coordinate sector and $\tilde{\kappa}$-deformed momenta \cite{60a}.
\newline
\indent ii) Snyder and Yang models are obtained by quantum group-theoretic
constructions, exploiting the $D=4$ and $D=5$ de Sitter algebras.
However, several Yang-like models, describing quantum deformed Lorentz-covariant
phase spaces were introduced by direct algebraic methods as well, based e.g.~on
the use of Jacobi identities (see e.g. \cite{20}, \cite{MM}, \cite{MM2211}).
\newline
\indent iii) The relativistic Snyder and Yang models are described algebraically in an equivalent way by $D=4$ and $D=5$ dS algebras. It is interesting to ask the fate of this equivalence if we consider \textit{quantum} dS algebras as Hopf algebras with \textit{nonprimitive} coalgebra sector. In such a case we can introduce the corresponding quantum-deformed Snyder models if the quantum Lorentz algebra is (in the Hopf-algebraic sense) the quantum subalgebra of properly chosen quantum dS algebras (for the choice of such quantum dS algebras, see e.g.~\cite{61a}).
\newline
\indent iv) In the Hopf-algebraic framework of quantum groups the generalized quantum
phase spaces can be obtained as the Heisenberg double algebra $\mathcal{H}=%
\mathbb{H}\rtimes\tilde{\mathbb{H}}$ (see e.g. \cite{62a,62b,31}), where $\mathbb{H}$ describes
quantum-deformed algebra with Hopf symmetries, $\tilde{\mathbb{H}}$ is the 
quantum Hopf group dual (in Hopf sense) to $\mathbb{H}$, and $\rtimes$ represents the so-called smash product
(see e.g. \cite{ABAP}). In such a scheme the Planck constant $\hbar$ appears as introduced in the Hopfian dualization procedure
\footnote{The Heisenberg double algebra for extended $D=4$ Snyder model has been
explicitly calculated in \cite{Meljanac:2021iyk}.}.

\section*{Acknowledgements}
We would like to thank M. Woronowicz for valuable remarks.
J.~Lukierski and A.  Pacho\l ~were supported by Polish  NCN grant 2017/27/B/ST2/01902.
A. Pacho\l \ thanks the Institute of Theoretical Physics University of Wroc\l aw for the hospitality during the scientific visit.
S. Mignemi and A. Pacho\l\  acknowledge the support of the European Cooperation in Science and Technology COST Action CA18108. 
S. Mignemi acknowledges support
of Gruppo Nazionale di Fisica Matematica.

\end{document}